\documentclass{article}

\usepackage{fourier}
\usepackage{amsmath}
\usepackage{amssymb}
\usepackage{soul}
\usepackage{authblk}
\usepackage{titlesec}
\titlespacing*{\section}{0pt}{4ex plus 1ex minus .2ex}{3ex plus .2ex}
\titlespacing*{\subsection}{0pt}{2.7ex plus 1ex minus .2ex}{2ex plus .2ex}
\usepackage[dvipsnames,table]{xcolor}
\colorlet{excolor}{blue!45!green!45!black}
\colorlet{rdfcolor}{black} 
\colorlet{shadecolor}{black!10}
\colorlet{goodgreen}{green!70!black}
\definecolor{pencolor}{HTML}{ADDFAD}
\definecolor{existencecolor}{HTML}{A1CAF1} 
\definecolor{narrativecolor}{HTML}{FAE7B5}
\definecolor{actioncolor}{HTML}{d9d9d9}
\definecolor{questioncolor}{HTML}{FFA07A}
\definecolor{questionscolor}{HTML}{FFA07A}
\definecolor{datacolor}{HTML}{D2B48C}
\definecolor{pertainscolor}{rgb}{0.56, 0.27, 0.52}
\definecolor{answerscolor}{rgb}{1.0, 0.75, 0.0} 
\definecolor{nuancescolor}{rgb}{0.9, 0.17, 0.31} 
\definecolor{detailscolor}{rgb}{1.0, 0.75, 0.0} 
\definecolor{instantiatescolor}{rgb}{1.0, 0.0, 0.5}
\definecolor{equatescolor}{rgb}{0.19, 0.55, 0.91} 
\colorlet{relatecolor}{black}
\usepackage{ragged2e}
\usepackage{amssymb}

\usepackage[includeheadfoot,margin=0.8in]{geometry}
\usepackage{lastpage}
\usepackage{fancyhdr}
\usepackage{graphicx}
\graphicspath{ {.} }

\newcommand{\mmmtype}[2][orange!60!black]{\textcolor{#1}{{\tt #2}}}

\def\poks{peices of knowledge}
\def\pok{peice of knowledge}
\def\pok{piece of knowledge}
\def\poks{pieces of knowledge}
\def\Poks{Pieces of knowledge}

\def\KCV{knowledge consumer view}
\def\KPV{knowledge producer view}
\def\KC{knowledge consumption}
\def\AC{attention consumption}
\def\KP{knowledge production}

\def\kcr{knowledge consumer}
\def\Kcr{Knowledge consumer}

\def\acr{attention consumer}

\def\kpr{knowledge producer}
\def\Kpr{Knowledge producer}
\def\CAR{collective attention resource}
\def\HTOTC{Hardin's TOTC}
\def\TOTC{TOTC}
\def\Leties{Local epistemic territories}
\def\lety{local epistemic territory}
\def\leties{local epistemic territories}

\newcommand{\mmmmark}[1]{\textcolor{purple!60!black}{{\tt #1}}}

\usepackage{caption}
\newcommand{\mmmtag}[1]{\textrm{$@$#1}}

\newcommand{\rdftag}[1]{\mmmtag{\textcolor{rdfcolor}{{#1}}}}

\usepackage{float}
\usepackage{enumitem}
\usepackage{hyperref}
\definecolor{links}{HTML}{288e47}
\hypersetup{colorlinks=true,urlcolor=links,urlbordercolor=links,linkcolor=goodgreen}

\usepackage{tikz}
\usetikzlibrary{shapes,snakes,shadows,arrows}
\usetikzlibrary{positioning}
\usetikzlibrary{calc,fit,matrix}
\usetikzlibrary{decorations.text,decorations.markings}

\tikzset{   
    existence/.style={fill=existencecolor, ellipse,align=center}, 
    question/.style={rectangle,fill=questioncolor,align=center},
    narrative/.style={fill=narrativecolor, rectangle,align=flush left}, 
    action/.style={rectangle,fill=actioncolor,align=center},
    datavalue/.style={fill=datacolor, rectangle,align=flush left}, 
    pen/.style={fill=pencolor, rectangle,align=flush left}, 
    edge/.style={font=\scriptsize,ultra thick,align=center,text=black},
    pertains/.style={edge, -latex, pertainscolor, text=black},
    relatesTo/.style={edge, -latex},
    answers/.style={edge, -latex, answerscolor,text=black},
    nuances/.style={edge, -latex, nuancescolor,text=black},
    details/.style={edge, -latex, detailscolor,text=black,font=\scriptsize},
    questions/.style={edge, -latex, questionscolor,text=black},
    relate/.style={edge, -, black},
    instantiates/.style={edge, -latex, instantiatescolor,text=black,font=\scriptsize},
    differsfrom/.style={edge, latex-latex, ForestGreen,text=black},
    pennedin/.style={edge, -latex, pencolor,text=black},
    equates/.style={edge, latex-latex, equatescolor,text=black},
    edgelabel/.style={font=\scriptsize,sloped},
    metadata/.style={draw=none,fill=none,anchor=south,inner sep = 0pt,font=\scriptsize},      
}

\title{\vspace{-1cm}A Knowledge Producer's View on the Knowledge Commons}
\author[1]{M. Noual}

\affil[1]{Université Paris-Saclay, CEA, List, Palaiseau, France}

\date{2023}

\begin{document}
\maketitle

Hardin \cite{Hardin} introduced the notorious concept of "\textit{tragedy of the commons}" (\TOTC). Worrying about the consequences of human overpopulation on the planet, he discussed what I will  refer here to as "\textit{hard problems}": problems with no technical solutions, that can only be addressed by way of an evolving morality. Hardin's \TOTC{} predicts that \textit{the hard problem of human population growth directly implies a hard problem of overuse or pollution of the commons}.
Here, I focus on the \textit{knowledge} commons. I argue that it is not clear that the \TOTC{} applies to the knowledge commons, for reasons similar to those that have protected agrarian commons for centuries. Even if the knowledge commons satisfies the \TOTC{}'s necessary conditions, it is not clear that the ensuing problems are \textit{hard}.
\smallskip

In section \ref{intro}, I present the point of view that I propose to take on the knowledge commons. It emphasises the notion of \textit{human attention}.  I also discuss conditions under which the \TOTC{} applies to the knowledge commons.  Section \ref{MMM} introduces the main principles of a technical solution called the "MMM" \cite{article} for structuring a digital knowledge record.
Relying on these technically implementable principles, section \ref{solutions} proposes solutions to push back and narrow down the hardness of familiar problems related with the knowledge commons \cite{simon1971designing,bawden2009dark}.
\smallskip

As will be justified below, in this paper, I don't distinguish between knowledge and information. 

\section{Tragedy or not?}
\label{intro}

The \TOTC{} is usually considered not to apply to the knowledge commons. In this section I argue that  there is a perspective on the knowledge commons from which the \TOTC{} does apply.
To predict the ineluctable overuse of a commons, Hardin made a number of assumptions. 
Primarily, he assumed consumption of the commons was \textit{rivalrous} and \textit{non-excludable}.\smallskip


Rivalry is due to the consumed resource being \textit{substractive}. This is usually not considered to be the case with knowledge 
\cite{hess2007understanding,rose1986comedy}:
 if I use a piece of knowledge, you can also still use that same piece of knowledge. 
Knowledge consumption isn't rivalrous and the conditions for \HTOTC{} 
aren't satisfied. 
There is no risk of overusing knowledge. Arguably it may even be the case that the more there are other individuals consuming knowledge the more educated society is and the more probable  it is for each individual to consume knowledge. 
This is a \textbf{knowledge consumer view} 
on the knowledge commons.
It focuses on the {consumption} of a non-material non-substractive resource, namely knowledge, assumed to have already been produced. With this view, production of the resource (e.g. by Fox News or by academic research) and also maintenance (e.g. fact-checking) are dissociated from consumption. They can be centralised.
The skills and means for emission and maintenance
can be entrusted to different entities\footnotemark{} than those who consume the information.~
\footnotetext{e.g. Google -- Google's guidelines \cite{googleguidelines} define at length what good information is under the hypothesis that good information is measured in terms of a fit between a document and a user's expectancies for this document. The webpage of a fascist conspiracist comic can be deemed of excellent quality if it is labelled as containing fascist, comic content. Serving this page to users asking for fascist comic content will bring about customer satisfaction, which is the purpose of "eyeball selling" digital businesses like Google \cite{evans2008economics,addison2007business}. 
However, arguably, good informing is a different activity from bringing about customer satisfaction. The Google founders themselves originally noted the incompatibility between good informing and eyeball selling \cite{googlewhitepaper}.
}
The reality of the so-called information age challenges this view. Informational resources are mass-produced. Centralised maintenance is not realistic. Institutional fact-checking is conspicuously failing at making the Web 
safe while Wikipedia continues to demonstrate the comparative success of distributed informational curation.
An alternative \textbf{knowledge producer view} is necessary to account for the {production of} the {knowledge} good,
 upstream from {knowledge consumption}.~
\smallskip

I assume that all \textbf{pieces of knowledge} (unlike thoughts) are produced with an intention for them to be consumed. With this assumption and the \KPV, production and maintenance are indissociable from consumption. \smallskip

Knowledge itself isn't subtractive, but the means of sharing it can be. If you borrow the  library's only book on Category Theory, then I can't borrow that book simultaneously. Digital technology mitigates the substractiveness of classical means of sharing knowledge \cite{tech}. 
The replenishabilty of network bandwidth  facilitates the replenishing of the potential for \KC{}.
The replicability of digital files facilitates the repetition of \KC{} experiences by multiple \kcr{s}. 
It would be a mistake however to confuse knowledge with knowledge media. Just like the book isn't the knowledge it conveys, radio frequency, network bandwidth, digital files are not the knowledge they convey.
The library book can serve as a door wedge. Its digital version can be used as part of a digital art piece. 
%
I propose to  relate \KC{}  with \textit{understanding and assimilating knowledge} rather than with \textit{accessing the (digital) resources that convey the knowledge}. 
The relevant bandwidth to consider now is that of the human mind: \textbf{attention}, or readiness to knowledge consumption
\cite{davenport2001attention}. Attention is  also replenishable, and arguably, non-capitalisable (present mental bandwidth can't be saved in order to have more bandwidth in the future). 
Considering attention reintroduces  the  question of substractiveness  \cite{davenport2001attention,injustice}.
If all my present attention is consumed in the intellectual effort to understand Category Theory, my mind is presently unavailable to learn about {AI transformers}.  
In 2004, P. Le Lay, then president of a French TV channel, notoriously said that what the TV channel was selling to Coca-Cola was "{available brain time}" \cite{cerveaudispo}. He specified that for an advertisement to be perceived, the viewer's brain must be available,  and that the purpose of the broadcasted TV content  is to prepare the brain for the reception of advertisement messages. Devenport and Beck introduce the notion of "attention economy" in which attention is a commodity, sometimes even a currency \cite{davenport2001attention}.  
Meanwhile, the digital marketing industry supports the "eyeball economy" where the good being sold by  media businesses is not digital information (that is provided for free to consumers) but consumers' "eyeballs", i.e., their attention
\cite{evans2008economics,addison2007business}. 
H. Simons said: "\textit{In an information-rich world, the wealth of information means a dearth of something else: a scarcity of whatever it is that information consumes. What information consumes is rather obvious: it consumes the attention of its recipients. Hence a wealth of information creates a poverty of attention and a need to allocate that attention efficiently among the overabundance of information sources that might consume it.}" \cite{simon1971designing}
\medskip

While there is no rivalry in knowledge consumption, there is rivalry to be expected in the consumption of the collective attention resource \cite{lorenz2019accelerating,mocanu2015collective,wu2007novelty}:
(1) epistemic \textbf{rivalry between \poks{}} and (2) (economic) \textbf{rivalry between \kpr{s}}. 
First, prior consumption of one specific piece or type of information  may affect the future consumption of another  \cite{newstead1992source,mocanu2015collective}. 
An individual's assimilation  of the theological argument may affect their  capacity to understand processes involved in biological evolution, and vice versa. 
The prevalence of an understanding of the theological argument in a community may affect that community's capacity to process biological knowledge, and vice versa.  
Of course not all \poks{} are in competition with one another. Some  synergise. One first \pok{} may make a second \pok{} easier to consume. 
A step further would be to consider a notion of "\textbf{epistemic pollution}" of information spaces. Just like air pollution degrades individual humans' experience of the air, epistemic pollution degrades their experience of  information. It consumes consumers' attention and their ability to be well-informed. A notion of epistemic pollution  would require a notion  of \textit{epistemic purity} which could be defined in different ways reflecting different biases on 
what pure/polluted knowledge is considered to be. 
From lack of a need to compare the qualities of different \poks{} in terms of a single informational quality, the present article doesn't need to formalise a notion of epistemic pollution. For the same reason it needn't distinguish between \textit{information} and \textit{knowledge}. I use the terms  interchangeably here\footnote{Mocanu et al. \cite{mocanu2015collective} provide an additional reason not to make a formal distinction between different qualities of information: "\textit{attention patterns when faced with various contents are similar despite the different qualitative nature of the information}".}.
Rivaly between \poks{}  competing for attention, makes \KP{} rivalrous for \kpr{s}.
If I produce a \pok{} that catches your attention,
I may have affected your ability to pay attention to someone else's work. \Kpr{s}, like/including the media industry's advertisers, compete for knowledge consumers' attention.
If the milk industry has already convinced the population that milk is scientifically proven to be good for your health, the scientific community may struggle to get  the nuancing \poks{} it produces  across.
\smallskip

To meet the conditions of the \TOTC{}, \KP{}, or more precisely \AC{},   needs to be "\textbf{non-excludable}"  in addition to rivalrous.
\textit{Physical} information media (e.g. printed newspapers) preserve a form of informational freedom or "intellectual privacy" \cite{infoprivacy}.  
They tend  to be slower and fewer at capturing our attention than digital media (cf smartphone notifications).
The ubiquity of information and communication technology makes it difficult to exclude \kpr{s} from tapping into the \CAR{} \cite{onlinebrain}. 
The non-excludability condition is satisfied, and the knowledge commons is subjected to the \TOTC{} after all.

\section{The Technical MMM Proposal}
\label{MMM}

The conditions of the \TOTC{} being satisfied, Hardin predicts  a problem of scale with no technical solution. 
Modern information and communication technology opens consumption  of the substractive collective attention resource to all. 
The size of the human population is bound to bring the knowledge commons to its ruin as 
the numerous \kpr{s}  acting in their own interest are bound to 
deplete the resource. 
Humans will then be deprived of the attention necessary to process information \cite{davenport2001attention} and the hard problem of overpopulation will become a hard problem of overly stupid population. 
To check Hardin's prediction, it remains to check the \textit{hardness} of problems involved. In this section I present a technical solution that invalidates Hardin's prediction by pushing back and narrowing down the need for morality to solve the issues of the knowledge commons (especially the need for a fresh new morality that wouldn't already exist).~
\smallskip

This section lists the main ingredients of the MMM proposal which is detailed  in \cite{article}. The idea of the proposal is to support the knowledge commons by organically materialising a "backbone" for it using the "MMM format"  (cf \S\ref{datamodel}). 

\subsection{Pieces of Knowledge}
\begin{figure}[H]
    \centering
    \begin{tikzpicture}
        \node[question]   (qsky)  {What colour is the sky?};
        \node[existence,left=4cm of qsky] (blue) {Blue};
        \node[question,above=0.3cm of qsky] (yesnoquest) {Is the sky is blue?};
        \node[narrative,below=1.2cm of qsky] (narrSkyBlue) {The sky is blue.};
        \node[existence,above left=0.5cm and 1.5cm of blue] (tobeblue) {To be blue};
        \node[existence,below left=1.2cm and 1.3cm of blue,inner sep = 0pt] (daytime) {the colour of a \\cloudless daytime  sky};
\node[existence,above right=0.5cm and 2.5cm of blue,inner sep = 1pt] (turquoise) {Turquoise};

\node[existence,left=4cm of blue] (bleu) {bleu};
\node[existence,below right=1.3cm and 2.5cm of blue] (colour) {color};
\path [draw,edge] (blue) to [bend right,in=-180,edgelabel] node (relate1){\colorbox{white}{\mmmtype[black]{relate}}} (tobeblue);
        \path [draw,answers] (narrSkyBlue) to [bend right,out=-90,in=-180,edgelabel] node (narranswer){\colorbox{white}{\mmmtype[answerscolor]{answers}}} (qsky);
        \path [draw,answers] (narrSkyBlue) to [bend right,out=-90,in=-90,edgelabel] node[yshift=-4pt] (narranswer2){\colorbox{white}{\mmmtype[answerscolor]{answers}}\\yes} (yesnoquest);
        \draw[answers] (blue) -- (qsky)  node[edgelabel,pos=0.4,yshift=0pt,fill=white] (blueanswer) {\mmmtype[answerscolor]{answers}} ;
        %
        \draw[instantiates] (blue) -- (colour)  node[edgelabel,midway,yshift=-8pt] (isardftype){\colorbox{white}{\mmmtype[instantiatescolor]{instantiates}}\\is a\\\rdftag{rdf:type}} ;
        %
        \draw[pertains] (daytime) -- (blue)  node[edgelabel,midway,yshift=-3pt] (bluedefinition) {\colorbox{white}{\mmmtype[pertainscolor]{details}}\\definition} ;
        \path [draw,differsfrom] (blue) to [bend left,out=70,in=145,edgelabel] node[yshift=-9pt,pos=0.6] {\colorbox{white}{\mmmtype[ForestGreen]{differsFrom}}\\add a bit of green $\rightarrow$\\$\leftarrow$ remove some green} (turquoise.west);

         \draw[equates] (bleu) -- (blue)  node[edgelabel,midway,yshift=-3pt] (bluebleu) {\colorbox{white}{\mmmtype[equatescolor]{equates}}\\language translation FR $\leftrightarrow$ EN} ;
    \end{tikzpicture}
    \caption{The MMM format defines different \textbf{types} of \poks{}. Some are represented in the figure above as \textbf{nodes} of a graph, others as \textbf{edges} connecting two nodes. Any \pok{} documented in MMM format must be assigned a type. In this figure, yellow rectangle nodes represent \poks{} of type "\mmmtype{narrative}" which is the \textbf{default type}. Orange rectangles and blue ovals respectively represent \poks{} of type "\mmmtype{question}" and  "\mmmtype{existence}". Yellow arrows represent \poks{} of type "\mmmtype[answerscolor]{answers}". 
    \textit{etc}. The visual choices made in this figure are arbitrary. MMM formatted information doesn't even need to be graphically represented
    \cite{MMMJSON}.
    MMM edge types (e.g. \mmmtype[equatescolor]{equates}, \mmmtype[ForestGreen]{differsFrom}, \mmmtype[instantiatescolor]{instantiates}, \mmmtype[pertainscolor]{details}, \mmmtype[nuancescolor]{nuances}, \mmmtype[questionscolor]{questions}), not all of which are illustrated above, have loose semantics which can be specified using an edge \textbf{label} -- e.g.  the epistemic relation conveyed by the \mmmtype[pertainscolor]{details} edge is specified by the label "definition".
        }
    \label{fig-examples}\label{MMMpres}
\end{figure}
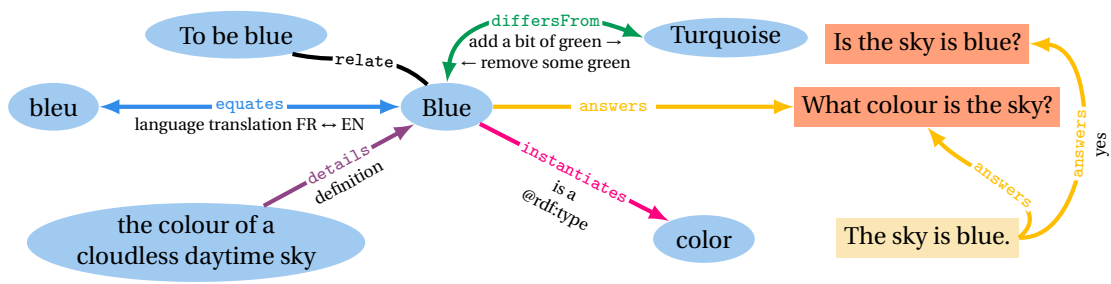

In the MMM proposal \cite{article}, the atomic informational unit considered is 
a \pok{}. Traditional documents (e.g. articles) usually contain multiple \poks{} and  sometimes overlap (one \pok{} appears in multiple documents).  Like the Semantic Web proposal, the MMM proposal departs from mainstream information technology approaches: it supports a finer granularity of information than what is supported by traditional documents.
The MMM format's (see next paragraph) definition of \poks{} however, 
is significantly looser than the SW's. Individual questions, in particular,  are typical  MMM \poks{} -- cf Fig. \ref{MMMpres}.

\subsection{A Common Documentation Format}\label{datamodel}

A comparison with the academic journal article format (AJAF) can help understand the MMM format.  
The AJAF is a common documentation format for academic researchers' work.
It imposes  a loose structure: 
documented knowledge is organised into standard types of "information containers" such as: \textit{title},   \textit{abstract},  \textit{introduction}, \textit{numbered section 2.3}, \textit{etc}. The \textit{introduction} of an AJAF document for instance can contain multiple \poks{}.
The MMM format is also a generic documentation format that imposes a loose structure. Unlike information containers of the AJAF, those of the MMM format are meant to contain a single \pok{}. The MMM format defines a small set of generic \textbf{types} of information containers which are presented in Figure \ref{MMMpres}.
Documenting notes in MMM format means decomposing them into a network of typed \poks{}. An especially important way of documenting MMM content is to \textbf{annotate} pre-existing MMM content. {Annotating} a  \pok{} $k$ (e.g. nuancing, questioning, detailing it) means documenting other \poks{}  and linking them to $k$ using the appropriate, predefined types of MMM edges (e.g. \mmmtype[nuancescolor]{nuances}, \mmmtype[questionscolor]{questions}, \mmmtype[pertainscolor]{details}). 
\smallskip

MMM formatted content can be saved in a human friendly MMM-JSON format \cite{MMMJSON}.
Existing software (e.g. documentation, publication, communication tools) may be enhanced with MMM formatting features, and various  new tools may be developed to support and to customise MMM editing and navigating experiences. 
An individual's access to the MMM formatted knowledge that they have produced and gathered is however not  dependent on any such tool. 

\subsection{Local Epistemic Territories}\label{terr}\label{LETs}
\begin{figure}[H]
    \includegraphics[scale=0.31, trim=0 4cm 0 4cm, clip]{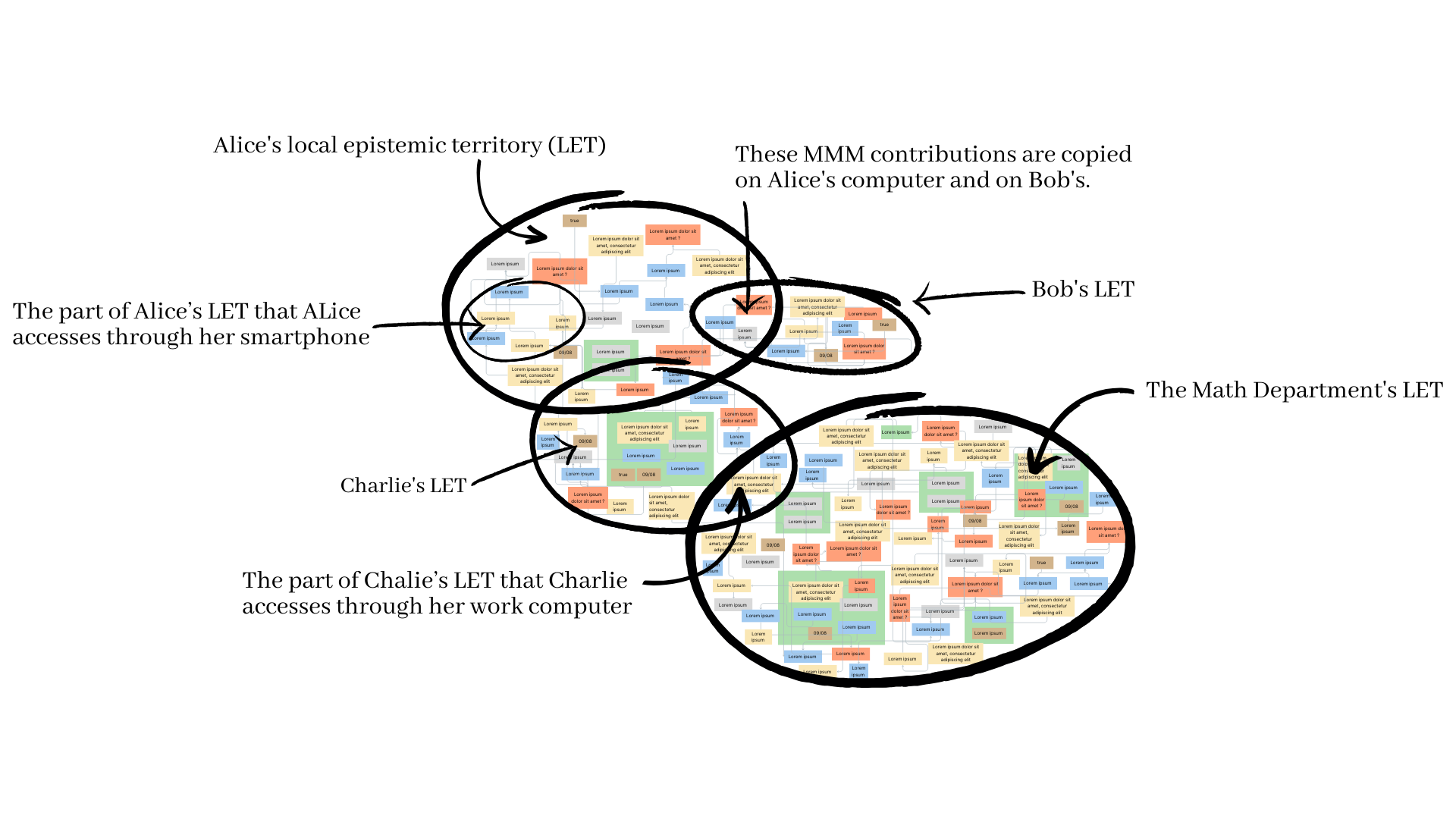}
    \centering
    \caption{The MMM  is a collection of overlapping local epistemic territories. 
    }
    \label{figLETs}
\end{figure}

A set of MMM \poks{} is called an \textbf{epistemic territory} or \textbf{land} \cite{article}. 
An individual's \textbf{local epistemic territory}
is  the set of \poks{} that she is acquainted with --  the area covering the extent of what she knows, cf Figure \ref{figLETs}. The \lety{} of a individual is  stored on  local machines that she owns and/or on remote servers that she controls. An individual has full agency of her territory. She selects the \poks{} that are added to it, and those that are deleted from it. 
Individuals can form  \textbf{communities} of interest together. Communities can equip themselves with servers to store the parts of their members' \leties{} that are of common interest. Community servers materialise the common epistemic grounds that are the community's raison d'être\footnote{They also can serve as backup storage for community members.}. 
Local machines and servers on which epistemic territories are stored, can be  owned by private or public entities. Epistemic landowners and/or tenants (community members) define the regulations that apply to their territories (cf \S\ref{gatekeeping}): who is entitled to access the \poks{} on their land, and under what conditions is a new \pok{} added to the land by a \kpr{}.
\smallskip

Formally, \textbf{the MMM (Mutual Mutable Medium)} denotes the reunion of all \leties{}, i.e. the set of all \poks{} documented in MMM format. As discussed below in \S\ref{distrib}, the MMM  is primarily a theoretical concept\footnote{Although the MMM proposal envisions MMM crawlers that leverage epistemic bridges across local MMM territories to gain hindsight on the global epistemic landscape.}.  In practice, only 
\textit{local} territories  of landowners (individuals or well-defined communities) are materialised.

\subsection{Organic Distribution}\label{distrib}
The classical \kcr{} view naturally sees  the knowledge commons as a global collection $\mathcal{C}$ of \poks{}\footnote{In the MMM solution, $\mathcal{C}$ would be the global MMM land, the reunion of all local epistemic territories.}. A noble aim is  to democratise access to  $\mathcal{C}$, make it wholly available to all \kcr{s}, typically by  \textit{decentralising} $\mathcal{C}$. 
$\mathcal{C}$ is  divided into chunks of data which are distributed to multiple storage locations. Some redundancy is  strategically implemented to ensure that even when one of the storage locations is down or disconnected, \kcr{s} all still equally  have access to the entirety of $\mathcal{C}$ at any point in time.
With the \KPV{}, the relevance of the concept of a single global collection of \poks{} is secondary. Primarily what matters is the local production of knowledge. The \KPV{}  acknowledges that knowledge is not centralised \textit{in the first place}e. Its processing  is contextual, its production  is localised, and its documentation is collective and thereby organically distributed \textit{from the start}. 
Even public knowledge (e.g. results from academic research) starts as private notes, produced by \kpr{s} whose work requires  some {intellectual privacy} and focus. 
Arguably, no single \kcr{} needs access to the entire knowledge record at any given point in time.
So it is not clear that centralisation of knowledge is necessary in the first place. And without some initial centralisation, decentralisation is unnecessary (and impossible). 
More important than democratisation of \textit{access} to knowledge is democratisation of means of getting messages across, especially getting them across to the right (ready) \kcr{s} equipped with the required amount of attention to process the information.

\subsection{Sharing and Publishing}\label{share}

In the MMM system \cite{article}, a \kpr{} can share a \pok{} $k$ from her \lety{} with other \kcr{s} of her choice who, if they accept $k$, include a copy of $k$ in their own \lety{}. 
Before sharing $k$, the \kpr{} can mark $k$ as "\mmmmark{public}". The \mmmmark{public} mark means "gifted to the public domain" and is intended  to be irrevocable. 
The author of a  \mmmmark{public} \pok{} $k$ is as much/little the owner of $k$ as anyone else who has a copy of it in their local territory. The only impact the author or anyone else can have on $k$ once $k$ has been published,
is to annotate  $k$ (nuance it, detail it, link it to a reformulation of $k$ \textit{etc}) in order to sway \kcr{s'} appreciation of  $k$ without directly modifying  $k$. People who have a copy of $k$ in their \lety{} can abstain from sharing $k$ and delete it. The \mmmmark{public} mark is irrevocable but the persistence of $k$ isn't. 
It reflects the intention of a \kpr{} but the reality may be different. Even if $k$ hasn't been marked as \mmmmark{public}, the consumption of $k$ may become a public matter if  no-one is excluded from it.
And conversely, if a 
 \mmmmark{public} \pok{} $k$ is filtered out of all public epistemic territories (cf \S\ref{grounds}), it can  only be found on private land and depends on private landowners to persist. 

\subsection{Epistemic Relations, Aggregation and Glue} 
\label{agg}

In the MMM proposal, links between \poks{}, namely MMM edges, are themselves \poks{} (e.g. the relation between Newtonian mechanics and Lagrangian mechanics). This means that  they can be  {annotated} (nuanced questioned, detailed, \textit{etc}) like any other \pok{}. 
MMM edges convey epistemic relations and play a central role in the MMM solution. 
\Poks{} linked by an edge or by a short path (sequence of consecutive edges) are said to be \textbf{epistemically close} -- e.g. all \poks{} illustrated in Fig. \ref{MMMpres}. 
The MMM proposal like the Semantic Web proposal is designed to facilitate the collective documentation of links between atomic informational resources. 
\Poks{} in the MMM have unique identifiers so any \kpr{} who wants to annotate a \pok{} $k$ that someone else has documented can do it as long as they (or the MMM software editor they are using) know(s) the identifier of $k$.
A major difference between the MMM and the SW is that the SW supports \textit{semantic} interlinking only, while the MMM more loosely supports any form of \textit{epistemic} linking \cite{article}.
Different answers to the same \mmmtype{question} $Q$ provided by different \kpr{s}, don't necessarily agree (aren't necessarily \textit{semantically} close) but they are \textit{epistemically} close in that they serve a similar epistemic role which is to answer $Q$. 
They can all be linked to $Q$ using a MMM \mmmtype[answerscolor]{answers} link -- in which case we say that $Q$ and its answers  are \textbf{agreggated} in the same area of the MMM. An \textbf{area} of the MMM is a connected set of \poks{}.
As another example,  a claim is epistemically close   to the definition of a term it  implies. The two can be linked together using a \mmmtype[pertainscolor]{details} edge labelled "is involved in".  
The MMM format is designed to support the \textbf{aggregation} of epistemically close \poks{}, i.e. the documenting of edges (or short paths) between epistemically related  \poks{}, to the effect that these \poks{} find themselves in the same area of the MMM. 
Simons argues that an information processing system (IPS) will reduce the information overload problem rather than compound it only if it absorbs more information than it produces  \cite{simon1971designing}: 
"\textit{To be an attention conserver [\ldots], [an IPS] must be an information condenser.  
It is conventional to begin designing an IPS by considering the information it will supply. In an information-rich world, however, this is doing things backwards.
}"
Thus, the MMM proposal's primary focus in on condensing the knowledge record (the MMM) by facilitating the shortening of epistemic pathways and aggregation (see also \S\ref{redund} about redundancy management).
%
\label{glue}
A single MMM edge-typed \pok{} between two \poks{} $k_1$ and $k_2$ is sometimes enough to express the epistemic relation between  $k_1$ to $k_2$\footnote{This is to be contrasted with hyperlinks on the WWW and citations in academic papers. Those two types of links are epistemically shallow: in themselves they convey no information on how the two \poks{} (in two different documents) are related. Notably, hyperlinks have been incriminated in shallow media multitasking \cite{onlinebrain,internetcog}}. Sometimes multiple \poks{}  are needed. 
The notion of \textbf{epistemic glue} generalises the notion of direct epistemic relation conveyed by MMM edges.
Epistemic glue between  $k_1$ and  $k_2$ denotes  a set of \poks{} that together 
express an epistemic relation between $k_1$ and $k_2$. 
When $k_1$ and $k_2$ belong to different \leties{} $T_1$ and $T_2$, the epistemic glue between them is said to epistemically \textbf{bridge} territories $T_1$ and $T_2$.

\subsection{Redundancy Management} 
\label{redund}

Epistemic glue can serve {epistemic democracy} through the exploitation of a \textbf{\textit{good}} form of \textbf{redundancy}. 
Suppose $k_1$ and  $k_2$ are two very similar \poks{} expressing  
the same idea ${\cal I}$ in different words appealing to different people. 
Documented epistemic glue increases the chances that people who understand ${\cal I}$ through the $k_1$ wording become aware of the  knowledge documented as following from $k_2$. 
Epistemic glue can also  help identify cases of \textbf{\textit{bad} redundancy} where   $k_1$ and  $k_2$ are so similar that \textbf{merging} them will result in no \kcr{} loosing potential to consume knowledge, on the contrary\footnote{\Kcr{s} initially only acquainted with $k_1$ and not $k_2$  gain direct access to \poks{}  documented as following from $k_2$, as a result of merging  $k_1$ with  $k_2$.}. 
When a \pok{} $k$ is published, it is usually meant to be seen.
If $k$ is redundant with \poks{} already documented, not only may its production have been a waste of the author's attention, it may also in the future waste the attention of \kcr{s} who will examine it only to realise it isn't new. 
Provision of epistemic glue, because it enables aggregation and thereby favours \kpr{s'} awareness of the state of the art in their field, may help preempt this situation. The more glue is provided, the easier it may be to notice the epistemic proximity between documented \poks{}, and the easier it may be to deal appropriately with redundancy -- get rid of it and avoid adding more.  

\subsection{Wayfarer Exploration}
\label{wayfarer}

The classic "\textbf{parachutist approach}" to discovering knowledge entails a \kcr{} making a request (e.g. typing a term such as "transformers" in a search engine, or querying a librarian) and having a centralised entity or process (e.g. a search engine, a librarian) decide what knowledge to make the \kcr{} aware of as a result of her query. With this approach, it is usually enough for a \kcr{} to know how to utter, spell or {type}  the term "transformers" 
in order to be served, among other things, (technical) knowledge about modern deep learning transformer technology \cite{googletransformersearch}.
The \kcr{}'s readiness  to consume  and understand the knowledge is not meaningfully accounted for in the knowledge retrieval and  distribution service. 
The MMM proposal unlocks the possibility for a complementary, alternative  approach.
With the MMM based "\textbf{wayfarer approach}", the \kcr{}  "walks" step by step, \pok{} by \pok{}, starting from  an area of their own \lety{} -- \textit{i.e.} starting from \poks{} that they are already acquainted with -- and expanding their \lety{} as they discover new \poks{} linked to what they know. The \kcr{} who knows nothing about deep learning, doesn't access a technical \pok{} about transformers until they have found an epistemic pathway between what they already know and the concept of transformers. The path need not make the \kcr{} understand transformer technology. It may be a very short path. Perhaps all the \kcr{} needs to find out before accessing the material is that there is indeed a reason for them to care about transformers. 
A hybrid approach to search can rely on/force wayfarer exploration in order to serve search results in response to a search query. 
An underlying assumption of {the wayfarer approach} is 
that on a daily basis, a \kcr{} is more likely to want to access resources that have some {epistemic relation} with the contents of their  \lety{}.
There must be another reason for Bob wanting to access a contribution about transformers,  than his ability to type "transformers" in a search bar. 
This other reason may already be reflected as a path in the MMM between what Bob knows and what Bob is interested in knowing. 
Otherwise, the wayfarer based search puts Bob in a \kpr{} position as he works his way over to the target concept of transformers, creating new epistemic links that he understands and materialising a new path in the MMM. 
Arguably, the purpose of documenting knowledge in general, and more precisely the purpose of documenting a line of reasoning ${\cal L}$, is \textbf{attention reuse}: to let future \kcr{s} benefit from pioneer thinkers' efforts in paving a way that  makes ${\cal L}$ less attention consuming for them (assuming their epistemic starting point is similar to the pioneers').
If there is an \textit{epistemic} reason for a \kcr{'s} interest in transformers, documenting this reason can eventually benefit other \kcr{'s} by  increasing the relevance and reach of their own wayfarer explorations. Generally,  any technical solution or incentive that promotes the connectedness\footnotemark{} of the MMM graph potentially improves  \kcr{s} wayfarer's explorations of the MMM. 

\footnotetext{Enhanced connectedness is a positive externality of the local provision of epistemic glue by individual \kpr{s}. }

\subsection{Implantation and Visibility}\label{implant}
A \pok{} $k$ is said to be \textbf{well implanted} in the MMM if (1) MMM edges are  documented between $k$ and other \poks{} and (2) these edges convey epistemically rich relations. Arguably,
if $k$ is recorded without any reference to pre-existing knowledge, it is of little value to humanity, until someone is able  to link $k$. 
On the MMM, good implantation of $k$ translates into high \textbf{visibility} of $k$. The more (and the \textit{better}) links there are between $k$ and other \poks{}, the higher the chances another \kcr{} will come across $k$ following the links that lead to $k$ (cf \S\ref{wayfarer}). On the contrary, lack of linking, and lack of \textit{good} linking may be severely detrimental to a contribution's visibility since no (good) pathways lead to it and the chances of other \kcr{s} finding it are low.   Epistemically shallow  or ill-positioned contributions (e.g. an unlabelled \mmmtype{relate} edge, or a \mmmtype{narrative} wrongly linked by an \mmmtype[answerscolor]{answers} edge to a \mmmtype{question} it doesn't answer) are likely to get {red-flagged} and/or filtered out by \kcr{s} (cf  \S\ref{gatekeeping}, \S\ref{contimp}). 
Provision of epistemic glue, be it through implantation of new \poks{}, the annotating of old ones, or the deliberate bridging of \leties{}, is the main way of contributing knowledge to the MMM. Floating knowledge islands without ties to the global MMM are practically invisible because they are unreachable to the wayfarer. \medskip

We can now emphasise the role of epistemic glue. Glue gives information on information. {The more of it, the easier} we may expect  administrative decisions concerning the global record of knowledge  to be, the more stringent and systematic the management of redundancy can be, i.e. 
the more intelligible the record becomes.
Epistemic glue explicitly documented in MMM format can be  \textit{generative} of a knowledge commons: glue can substantiate both (i) the content of the commons and (ii) its structure,  spanning across \leties{} thereby allowing us to speak of a common overarching domain. Arguably,  epistemic glue is key to the safe scaling up of the knowledge commons: 
as more \kpr{s} contribute to it, more attention reuse enabling glue is documented. 

\subsection{Epistemic Topography and Shortsightedness}\label{topo}

The MMM format gives a graph like structure to \leties{}. \textbf{Epistemic measures} can be defined in terms of graph theoretic  properties. 
For instance, the depth (resp. utility) of a  \pok{} $k$ can  be measured by the length of paths  incoming (resp. outgoing) $k$. Less naive notions of depth and utility can take into account the types of MMM contributions involved (especially the types of edges).
Epistemic measures are a technical solution to the (hard) problem of evaluating information quality in more diverse and relevant terms than in traditional binary {true}/{false} knowledge qualifiers.  
And they may be used to visually represent epistemic territories as 3D landscapes to \kcr{s}.
\smallskip

\label{igno}
Arguably, \kcr{s} are entitled to  an amount of ignorance, disinterest, and misconception.
Scientific advancement would be impossible if scientists weren't allowed to grope their way to new knowledge. 
Good information does not exist without awareness of what one ignores and of the limits of the knowledge one has. Formal MMM based epistemic measures  can support \textit{safe} ignorance.   
\Kcr{s}, unequipped to process the technical details of a \pok{} $k$, can still be aware of the existence,  epistemic purpose and depth of $k$ by way of appropriate  epistemic measures. {Ad hominem arguments no longer are \kcr{s'} main way out of a requirement for an expertise they don't have.}  


\subsection{Non-findability and Gatekeeping}\label{nonfindable} 

With the \KCV{}, it may be tempting to simplify  epistemic democracy  to  universal access to all public knowledge (cf \S\ref{distrib}). With the \KPV{}, a finer notion of epistemic democracy is possible, accounting for the futility of distributing knowledge to individuals and communities that are not prepared for it, and for the importance of building bridges (cf \S\ref{glue}) across worldviews (\leties{}) in order to gently enhance worldview fluidity  and democratise epistemic readiness  without aggressively "talking to walls" in an attempt to brutally negate and replace inconvenient worldviews. 
The technical MMM solution  supports a form of subjective {"\textbf{non-findability}" or "\textbf{non-spontaneous findability}"} that departs from traditional enlightenment ideals: Bob's ability  to type "transformers" \textit{per se} no longer opens the way for him to any public material associated to the keyword "transformers"\footnotemark.
The actual state of Bob's \lety{}, which no-one needs to know about but Bob, decisively determines what \poks{} are non-findable to Bob. And Bob's epistemic wayfarer (i.e. learning) efforts
determine what \poks{} become non-spontaneously (effortfully) findable to him (i.e. how his \lety{} expands).
Note that if a \textbf{central entity} were to define what \poks{} are  non-findable to  Bob, that entity would have to know an uncomfortably huge amount of information on Bob and Bob's actual situation (probably no less than what Bob has) in order to determine what knowledge  Bob is able to process with the mindset he has today. The organic distribution  of the MMM (cf \S\ref{distrib}) avoids the hard problem of  defining non-findability and having a central entity implement it.\smallskip

\footnotetext{Bob can still be made aware \textit{from a distance} of the existence of technical material, say for LLM specialists,  through epistemic measures (cf \S\ref{igno}). Non-findability of a \pok{} $k$ concerns the \textit{consumption} of $k$ by Bob, not his \textit{awareness} of $k$. This is to help avoid unprepared \kcr{s}  misreading and misusing  $k$. }

\label{filter}\label{gatekeeping}

A \kcr{} can grow her \lety{} by adding copies of  \poks{} discovered while exploring the MMM (cf \S\ref{wayfarer}) or by accepting \poks{} shared with her by known \kpr{s} (cf \S\ref{share}).
The \poks{}  added to a \kcr{'s} \lety{} are the ones responsible for consuming her  attention.   
The wayfarer is a technical solution  to help a \kcr{} protect  her attention 
and  prevent attention-wasting \poks{}  from finding their way to her.   \textbf{Epistemic gatekeeping} is another which consists in 
systematically rejecting \poks{} based on how they measure in terms of well-defined epistemic properties (cf \S\ref{topo}).  
MMM based epistemic measures  open the possibility for \kcr{s} to fine-tune automatic filter mechanisms.
They can customize the conditions under which a \pok{} shared with them is automatically accepted as part of their \lety{} or rejected. For instance a \kcr{} might want to systematically ignore \mmmtype{narrative}s that no-one has yet deeply (cf \S\ref{topo}) supported, questioned or nuanced.
{Gatekeeping} operates at the finest level of epistemic granularity (that of \poks{} rather than that of documents and document authors). It  increases \kcr{s'}
capacity to exclude \poks{} and exclude authors from wasting their attention. Belonging to a topic $T$ of interest to Bob, or mentioning keywords statistically related to $T$ is no longer enough for a \pok{} $k$ to reach Bob (to be parachuted onto him). Bob's \lety{} must demonstrate a \textit{measurable} form of readiness towards $k$. 
\smallskip

\label{echochambers}

In the current document based world, facilitating the circumscription and gatekeeping of \leties{} 
would risk favouring \textbf{echo chambers}. Why can things be different in the MMM based world? Because of an  {interplay} between gatekeeping filters (\S\ref{gatekeeping}) and epistemic bridging/glueing (\S\ref{glue}). Filters  prevent \kcr{s} from seeing \poks{} they don't want to see. Epistemic glue (aggregation) 
makes any \kcr{} more likely to see all \poks{} that are  epistemically close to every \pok{} she takes interest in. 
This includes the \mmmtype{question} $Q$ that is answered
by the  belief $B$ she has,  as well as all documented alternative  answers $A$ to $Q$ which contradict $B$ and possibly nuance her worldview. 
Filters are based on \textit{epistemic} properties of \poks{} (cf \S\ref{topo})  rather than \textit{semantic} properties. This means that \poks{} can't be filtered out solely based  on what they mean. They can only be filtered out based on their position in the MMM, i.e. their relation to other \poks{}. 
Alice can easily filter out everything  documented as an answer to \mmmtype{question} $Q$. She can also invest some of her own attention in the formalisation of a measure of incompatibility of an answer
with her belief $B$ -- e.g. exploiting possible \mmmtype[ForestGreen]{differsFrom} links.
It is however not obvious how she may define general epistemic measures that reliably recognise all \poks{} in disagreement with her worldview. Aggregation 
makes the walls of  echo chambers   porous. 
Anyone at the margins of a closed epistemic community that is caught in an echo chamber, can play a liberating role if they import even slightly more nuanced \poks{} into the community's \lety{}.

\subsection{Biased Common Grounds}\label{grounds}

Discussing a particular \pok{} $k$ on one \lety{}, say Alice's, is generally not equivalent to discussing $k$ on any other \lety{}.
Let $T_E$ be the \lety{} of an entity $E$, either an individual like Alice or a community (e.g. conspiracy theorists  or the academic community). 
Suppose that  $T_E$ includes the \mmmtype{question} "\textit{How do RNA vaccines work?}" and an answer $A_1$ mentioning microchips and 5G technology. An alternative answer $A_2$ exists, mentioning surface proteins. $A_2$  is documented on a different entity's \lety{}, but not on $T_E$. $E$ is responsible for what \poks{} are accepted as part of $T_E$. $E$ has epistemic biases and interests. Because of that, $T_E$ does not provide neutral grounds to have a discussion on the risks of RNA vaccines. Indeed, answer $A_2$ not being represented on $T_E$, it can't be annotated and discussed on $T_E$.  \Leties{} are not equal in terms of the possibilities of discussion they support. Understanding a \pok{} $k$ on a \lety{} $T_E$ (by looking at the contextual knowledge surrounding $k$ in $T_E$) is also usually \textit{not} equivalent to understanding $k$ on any other \lety{}.
Some \poks{} critically relevant to $k$ might be absent from $T_E$.
Some annotations and questions might be impossible on $T_E$. 
If a \mmmtype{public} \pok{} $k$ is of value (e.g. a scientific publication that some scientists are  willing to keep a local copy of), then $k$ should be discussable on \textit{public} common grounds. To discuss $k$, one should not have to be invited on someone's private \lety{} where the owner's conversational rules must be respected  and his chosen gatekeeping applies\footnote{This departs from project Solid's notion of pods \cite{ruben}.}. 
Public institutions  have a key role to play in the MMM ecosystem: that of providing the most \textit{neutral} discussion grounds possible -- mitigating risk of systematic epistemic attention deficit  \cite{injustice} --  and ensuring these grounds are continually updated as the state of documented knowledge evolves.
It remains for  public institutions to decide what  \poks{} are of value (worth discussing) and what are not. This is similar to what any entity managing a local  communal epistemic territory needs to do, except that the purpose of the public entity's bylaws is not to support a common interest but resolutely to enable open public discussion of public knowledge.
Local public territories must not be restricted to a set of trustworthy  \poks{} such as peer-reviewed scientific publications. 
To avoid {the fact-checking pitfall}, information traditionally deemed of low quality (such as unsubstantiated or cryptic conspiracist contributions) should be included,
 not censored, cf \S\ref{contimp}. 

\subsection{Continual Improvement}\label{contimp}

It is impossible to produce new information without producing low quality information (cf the daily practice of scientific research)\footnote{This doesn't mean that (valuable) new information is necessarily produced when low quality information is produced! Contributions can be so shallow that the best way to deal with them is to shut them down (address them) early to discourage any repetition of them later. }. Many errors can decisively participate in the process of improving information. The contribution of a low quality \pok{} $k$ is a step in the process of improving the record: it calls for further contributions specifying what is wrong with $k$  and how to deal with it\footnote{If the $k$ is very shallow, there might not be anything else to do than to address it once and for all and ensure all repetitions of it funnel to the same area and aggregate, so as to take a minimum amount of space in the knowledge commons/in the \CAR{}.}.
$k$ should not be removed from the commons. Nor should its author be scorned\footnote{The MMM solution proposes to penalise \kpr{s} who repeat \poks{} not \kpr{s} who produce low quality \poks{}. It aims at protecting the attention of \kcr{s} without reducing the freedom of expression of \kpr{s}. }. A contribution of poor quality, just like any other contribution should be systematically exposed to the public eye, subjected to discussion and improvements, 
durably enough that concerned citizens learn from it\footnote{Arguably, it is not so much misinformation that is a problem on actual information spaces, than it is its diffusion \cite{mocanu2015collective}. Hindering the diffusion of misinformation requires telling it from information. An alternative  is to change the way it diffuses. The MMM solution, through aggregation, proposes to ensure (mis)information systematically diffuses with its {aggregated} "reading notes" (nuances \textit{etc}) glued to it. 
Another  possible factor at play 
suggested by \cite{mocanu2015collective} is the habituation of \kcr{s} to unsubstantiated claims. Systematic exposure to nuances and awareness of the  topography around a claim (cf \S\ref{topo}), including lack of nuance, are technical ways to mitigate this habituation.
}. Visibility of an error and how it has been addressed should be entertained as long as citizens  risk  repeating the error. 
On the MMM, annotations brought to $k$ (\mmmtype[nuancescolor]{nuances}, \mmmtype[pertainscolor]{details}, \mmmtype[questionscolor]{questions} \textit{etc}) reflect  the collective understanding and processing of $k$.  
Especially if $k$ is deemed low in terms of informational quality,  it is preferable to persist $k$ and its annotations 
for as long as the topic is hot. 
Annotations to $k$ improve the record of knowledge in that they make the record more helpful to future \kcr{s} considering $k$. They make the work of processing $k$ less attention-consuming since the work 
has already been done and documented in $k$'s annotations.
\smallskip

A traditional approach to ensuring the quality of a knowledge space such as the Web or scholarly communication is to hunt low quality content out of it and accumulate new higher quality \poks{} instead.  
The MMM solution supports an alternative approach 
focussed on \textbf{attention reuse} and on learning to deal with existing \poks{}.
\medskip

Mechanisms that enhance and exploit the connectedness of the MMM increase the chances that a \kcr{} can reuse the attention she spent consuming (understanding) a \pok{} $k$ in the past to consume a new \pok{} $k'$ glued to $k$. 
The MMM solution 
incentivises (cf \S\ref{implant})  authors to strip the MMM formatted expressions of their contributions down to what is really new. Aggregation makes repetition of atomic \poks{} pointless. Implantation-based visibility  also incentivises the \kpr{} of $k$
to make substantial 
reference to relevant pre-existing \poks{} $k'$ (whose visibility $k$ may inherit), so that the attention needed by \kcr{s} acquainted with $k'$ to process $k$ is reduced. 
\smallskip

\Kcr{s} whom deem a contribution $k$ to be of low quality can help improve the knowledge commons in several ways:
(1) if they are willing to spend some of their own attention on $k$, they can write and share a comment on $k$ (\mmmtype[questioncolor]{question it}, \mmmtype[nuancescolor]{nuance it}, \textit{etc}) making  the reason of their disapproval explicit and known, (2) if $k$ is ill positioned in the MMM (e.g. $k$ is linked by an \mmmtype[answerscolor]{answers} edge to a \mmmtype{question} that it clearly doesn't answer), they can "red-flag" $k$ without investing much attention in $k$ \cite{article}, (3) they can also refrain from sharing $k$, or on the contrary, (4)
if $k$ is thoroughly discussed (challenged and nuanced), they can widely share $k$ together with its annotations  to mitigate the risk of other \kcr{s} processing $k$ without its nuances.
\smallskip

I hypothesise that shallow unsubstantiated contributions (e.g. troll provocations) will "behave" differently on the MMM than reliable \poks{} (e.g. methodically produced and rigorously peer-reviewed scientific results): they won't be annotated in the same way. 
Formal epistemic measures (cf \S\ref{topo}) will provide a means to tell different qualities of \poks{} and to deal with them accordingly 
without resorting to censorship on the global MMM, while still allowing individual \kcr{s} to experience a finely filtered version of the knowledge commons.

\subsection{Activity based reward}
\label{precedence-authorship}\label{rewarding}

The MMM format requires every documented \pok{} to be assigned one or several "authorships". An authorship is given by a team of authors and a timestamp.
Two independent (teams of) \kpr{s} can be authors of the same \pok{}. Precedence of authorship is mostly disregarded. Who among Alice or Bob published resource  $k$ first in time doesn't matter. What matters is that  $k$ is published, that Alice's version of  $k$ and Bob's version of $k$ be identified and documented as identical as soon as possible, and that Alice and Bob be both rewarded appropriately for the effort they invested. 
MMM based epistemic measures (cf \S\ref{topo}) can help measure the epistemic value of a \pok{}. 
Perhaps more importantly, they can be used to evidence a \kpr{'s} characteristic expertise and its value to the knowledge commons. For instance, some \kpr{s} are good at formulating fundamental \mmmtype{questions} that prompt other \kpr{s} to provide new answers. Other \kpr{s}  improve the knowledge commons by contributing  bridges between epistemic fields. Measures can be defined to capture patterns in \kpr{s'} contributions. \Kpr{s} can thus be rewarded accordingly for their  proven producing competences 
rather than for their products \textit{per se}, i.e., rather than for their ability to consume collective attention. The MMM proposal's emphasis on epistemic implantation (cf \S\ref{implant}) and glue (cf \S\ref{glue}) supports rewarding of \kpr{s} for their participation in the 
continual improvement  of the knowledge record which involves mitigating the attention required to navigate the record in a way that is compatible with enhanced learning and  good informing of \kcr{s}.
The networked structure of the MMM also allows for a "\textbf{trickling reward} system" \cite{article}:   paths in the MMM graph can be leveraged to acknowledge the participation of "little hands" involved in a chain of \kpr{s} who all contributed knowledge that eventually led one successful \kpr{} among them to produce a \pok{} formally rewarded outside the MMM (e.g. by a prestigious publication or  prize).

\section{Solutions in Support of the Knowledge Commons}\label{solutions}

Section \ref{intro} concluded that from a \kpr{'s} perspective focussing on \AC{}, the rivarly and non-excludability conditions for \HTOTC{} are satisfied. 
Many authors have gathered evidence and arguments casting serious doubt on the  validity of the \TOTC{} in general \cite{rose1986comedy,OstromGov,Cox}. 
Here, with the \kpr{'s} view, the  
\TOTC{} instantiates to the following implication:
\textit{the hard problem of human overpopulation directly implies a hard problem of 
overuse of the knowledge commons} (\textit{by \kpr{s}}), that is, \textit{it implies a hard problem of knowledge overload}, or \textit{overuse/pollution of the collective attention resource}. 
I propose to re-examine now the validity of this implication, and in particular the \textit{hardness} of the problems it predicts, in the light of the technical propositions made in section \ref{MMM}.
\smallskip

Excess of (digital) information media (e.g. Web content) seams to be an actual problem
\cite{ho2001towards} which may indeed be aggravated by overpopulation, since modern information and communication technology democratised publication.  
It is not clear that excess of knowledge \textit{per se} is also a problem, nor that it is an \textit{actual} one
\cite{openquestions,park2023decline,ward2021people,kets2023posttruth,knudsen2022multiplying,schwenkenbecher2022we}. 
The democratisation of digital content production has yet to be matched with the democratisation of knowledge production. 
The possibility -- offered by technically supported epistemic aggregation   (cf \S\ref{agg}) -- of condensing digital content to its epistemic pith 
 is a first reason to doubt Hardin's implication. 
A second reason  is the following. As Hardin suggests, human overpopulation is ethically tricky to address. Hardin mentions how the bird population regulates itself: bad bird parents produce less viable chicks. Their chicks tend to die from bad parenting. Human society strives to save human babies from bad parenting. What works for birds doesn't for humans. However, while human ethics frowns upon throwing babies away, even when two copies of the same baby have been produced, {throwing \poks{} away} is a different matter, especially in case of duplicates (cf \S\ref{redund}). 
So even if overpopulation does imply overload of the knowledge commons, it is not clear that the \textit{hardness} of the problem of overpopulation translates into the hardness of the problem of overuse. 
\smallskip


It would indeed be ethically tricky to prevent some human parents from reproducing themselves. But a commons isn't necessarily a good that ethics require keeping open and unregulated for  all to enjoy like the right to reproduce. 
Hardin relies on the historical example of  \textbf{agrarian commons} to support his concept of \TOTC{}. 
As abundantly noted in the literature, agrarian commons were usually far from being open to all and unregulated  as Hardin assumed \cite{OstromGov,Cox,Netting}. Villagers entitled to use the communal land had to follow strict rules and limitations. 
The  non-excludability condition didn't apply so nor did the \TOTC{}.
Agrarian commons perdured for centuries. Some still exist~\cite{crosetti2021mountains} -- a situation, S.J.B. Cox  notes,  might have better been described as the "triumph of the commons" \cite{Cox}. 
The \TOTC{} can generally be avoided by avoiding
either one of the conditions under which it operates.
The substractive nature  of attention  is difficult to avoid. But the {non-excludability} of  its consumption isn't a fatality: non-excludability isn't intrinsic to attention. 
As mentioned in 
\S\ref{intro},  it follows mainly from highly efficient digital means of communication that let any \kpr{} reach a lot of \kcr{s}. An approach to making 
\AC{} excludable would be to bridle modern technology. 
An alternative is to use it for sharing  knowledge between well-defined, rigorously regulated 
epistemic territories (cf \S\ref{terr}). \linebreak
\vspace{-6pt}

Note that without excluding  anyone from the possibility of exercising their right of access to a good (e.g. reproduction, publication), \textbf{tools can be built} to facilitate some people's exercise of that right. 
Languages (e.g. French, mathematics, C) are common goods. Without formally excluding anyone from speaking French, French dictionaries make it easier for French people to speak more French and do nothing for non-French speakers. 
Of course the {indirect promotion of some people's enjoyment of a good} can in some cases pose ethical problems of unfair use. This depends on the good.  Arguably, it is to everyone's benefit that aircrafts occupying the international airspace  commons can only be operated by trained pilots rather than by any layperson who fancies piloting a plane. On the other hand,  social media democratically equips people with tools that amplify enjoyment  of the common human ability to communicate emotions of joy and indignation. I contend it is time to consider balancing the impact it has on society with
technology that amplifies the enjoyment of  the common human ability to carry out analytical reasoning. 

\subsection{Locality}\label{locality}

An essential feature of historical communal lands that Hardin overlooked  was {locality}.
Locality has been decisive in assigning communal tenure or not to a land. Communally tenured land is \textbf{not just any land}: it can be the wasteland privately owned by a local lord,  a remote land owned by a specific group of villagers, a forest land that is difficult to control individually  \cite{Cox,Netting}.
Because of their remote position in the mountains, certain uses of the alps for instance, were possible (like herding goats in the summer)  while other uses remained impossible (like growing crops for subsistence), \textit{even in the absence of any regulation}. 
Use of the land is primarily governed by: (\textit{i}) what kind of land it is, (\textit{ii}) who has tenure of it, and  (\textit{iii}) who owns\footnotemark{} it (cf \cite{OstromGov} for a more thorough analysis). 
\footnotetext{
  Historical communal lands could be private or public. Property regimes are not of primary interest to our discussion. 
  A more important question is whether the land is individually or communally tenured. 
  As history demonstrates for agrarian land, and as the MMM solution defines for epistemic land, both tenure regimes  are possible and can relevantly co-exist. 
  Communal agrarian land was generally not land that could in itself suffice to the activities of farmers. If there was communal tenured land there likely was individual tenured land. This lead to the seasonality rule discussed in \S\ref{seasonality}.
}
Of course, a landlord decides to which individual(s) the tenure of the land is granted ((\textit{iii}) influences (\textit{ii})), and under what conditions it is. 
 Activities of tenants on the land modify the properties of the land ((\textit{ii}) influences (\textit{i})). Conversely, properties of the land, including its geographical position influence who wants and can have tenure of it.  A Chilean fisherman in the 1570s was naturally excluded from claiming access to an Alpine pasture. A  parallel with the knowledge commons can be made: not everyone necessarily needs access to all resources (cf  \S\ref{distrib}). The relevance of considering the collection of all common   resources  \textit{as a whole} is clear neither in the case of physical resources nor in that of epistemic resources. 
Furthermore, similarly to physical land, as specified in \S\ref{grounds}, not all epistemic territories are equivalent. It is natural to expect different communities to rally around different epistemic grounds and perform different epistemic activities. 
A miniature model community may spontaneously cultivate a territory that is grounded in chemistry science because of  observations its members make about solvent properties and interactions. This isn't necessarily  because of gatekeeping. It may be because of  the territory's theme (miniature modelling)  and the curiosity and astuteness of miniature modellers. 
In general, different areas of a \kcr{'s}  \lety{} might matter differently to them. 
Loose gatekeeping (cf \S\ref{gatekeeping}) of the areas  that they consider to be  frivolous  (e.g. celebrities'choice of shoe style) might be enough. Outsourcing the gatekeeping of those areas 
to a central entity (e.g. a community manager) might be acceptable.  Appropriate gatekeeping of other areas (e.g. the individual's scientific research work) might require the individual's involvement. Some epistemic land (e.g. semi-mature mathematics research) might be better and/or more easily cultivated {communally} by a community of domain experts. Some might require to be contributed by a very small number of expert \kpr{s}. Some might be safely open to large numbers of \kcr{s}. Some might be better kept  private and confidential, \textit{etc}.

\subsection{Seasonality}\label{seasonality}

Netting \cite{Netting} reminds us of a simple historical seasonality rule   that still governs  the grazing of some communal alps in the summer: "\textit{No citizen can send more cows to the alp then they can feed during the winter with the harvest of their own hay meadows}". Bad farmers who disrespect this natural rule and overuse the commons, are likely to individually suffer the consequences. 
In the case of the knowledge commons, is such a straightforward rule possible? Is there a natural limit on the amount of \CAR{} we can expect reasonable competent \kpr{s} to (already) be (partly) incentivised to respect? To give a positive answer to this question, we must look for the conditions under which it is \textit{not} in the obvious immediate interest of a \kpr{} to consume more of the \CAR{} rather than less. Indeed, an approach to solving a hard problem is to identify users who already implement moral principles that mitigate the problem, and build a technical solution to enhance those users' influence. 
We have assumed that knowledge production is indissociable from an intent of knowledge sharing. 
However, we can still dissociate (1) the act of producing knowledge from (2) the act of sharing  the end product. 
Sharing knowledge consumes \CAR. 
Producing knowledge consumes the individual attention of the \kpr{(s)}. More precisely, the output of a new \pok{} requires the input of a limited number of pre-existing \poks{} and the attention of a small number of collaborating \kpr{s}. 
So a straightforward form of seasonality applies here too.  
Before they consume any collective attention in the sharing of their work, \kpr{s} must invest some of their own attention in the crafting of a \pok{} worth sharing. 
This  \KP{} seasonality rule is  spontaneously implemented by reasonable competent \kpr{s}   who refrain from communicating a contribution until they have spent enough of their own attention on it. 
The rule promotes a form of intellectual privacy: no-one needs to know what the \kpr{} is producing until the \kpr{} expertly deems their work to be ready.  \Kcr{s} are naturally excluded from consuming a \pok{} before its time. With the MMM solution, a \kpr{} starts by privately documenting her work in her own \lety{}. The pieces of her work are   individually released  when she decides. 

\subsection{Bad Farmers}

The seasonality rule applies to information \textit{relayers,} \textit{i.e.} individuals who participate in the diffusion of a \pok{}  that they have not "crafted" with their own minds (e.g. they retweet a message on Twitter). Information relayers are \kpr{s} themselves.
They produce copies of \poks{} that are susceptible of consuming the \CAR{}. Like first-hand \kpr{s}, relayers should be expected to invest some of their own attention into the information they relay. \textit{No individual should send more information into the information commons than they can invest attention in mastering with their own attentive intellect.} 
In economic terms, individuals who participate in the knowledge communication chain without investing a sufficient amount of their own attention into the information they relay are \textit{free-riders}. They benefit from collective attention  (their contributions are seen) but they underpay for it with their own attention.
Free-riding \acr{s}  are much worse than free-riding \kcr{s}  since as mentioned before, unlike \KC{}, \AC{} is substractive. 
Excessive free-riding risks depleting the \CAR{}, spending it on  shallow redundant information.
On the MMM, relaying a \pok{} $k$ consists in sharing the epistemic coordinates (identifier) $k$. It produces no new content and is limited by local gatekeeping. 
True free-riders cause the knowledge record to grow in a way that is wasteful of collective attention. 
Implantation (cf \S\ref{implant}) provides a partial  technical solution to free-riding. Rather than directly punish free-riders (whose behaviour might stem from a lack of epistemic education),
implantation disfavours their contributions and helps enforce the seasonality rule.
Free-riding relayers who don't pay enough attention to the way  their contributions epistemically relate / add to pre-existing knowledge\footnotemark{}
will find it hard to ensure the visibility of their contributions.
If they do manage, they are not free-riders. 
\footnotetext{Note that the MMM proposal leaves \kcr{s} free to "fast-consume" knowledge (without spending their own attention), and even free to leave it to external entities to  decide what knowledge they are fed. \Kcr{s} can even accept to be fed by an  external centralised entity (e.g. Google) with a primary agenda that departs from their good information  -- a "leach's agenda" of consuming the attention of the \kcr{} for their own interest (e.g. selling eyeballs to advertisers) -- as long as the \poks{} that \kcr{s} are fed land and remain in their own epistemic territories and aren't systematically passed on to the commons. A parallel can again be made with physical land. The landowner is free to enjoy their land and bring things on their land (e.g. pesticides) as long as they don't affect other people's enjoyment of the surrounding land. 
If a \kcr{} starts systematically relaying the \poks{} that they fast-consume, then they start acting as a free-rider consuming collective attention, mostly for the agenda of their favourite attention-leaching feeder.}
To illustrate this, suppose Alice produced a \pok{} $k$ which is now well implanted in an area $A$ of the MMM. Later Bob produces a very similar (possibly identical) \pok{} $k'$ (possibly even, Bob's $k'$ comes from Alice's $k$).
A cooperative behaviour from Bob means Bob endeavours to implant $k'$ in an area $A'$ of the MMM of interest to the same people as $k'$.
If there are several such areas then documenting $k'$ is an opportunity to document epistemic glue between them. 
 $k'$ is likely to end up close to $k$, and $A'$ is likely to overlap with $A$, even if Bob is initially unaware of the similarity between  $k$ and $k'$. 
The similarity between $k$ and $k'$, once noticed by the community of \kcr{s} is likely to bring about the documentation of epistemic glue   between $k$ and $k'$.  
If Bob is aware of Alice's $k$, he might fear that the precedence of Alice's contribution  
limits his chances of getting rewarded for $k'$. 
Uncooperatively, he might decide to entertain the illusion of the novelty of $k'$ for as long as possible.  To hinder and delay the identification of the similarity between $k$ and $k'$, Bob needs to implant $k'$ in an area $A'$ that is as far away as possible from area $A$.
This means depriving $k'$ from inheriting the visibility of $A$.
To ensure he is still rewarded, Bob must  choose $A'$ well and implant $k'$  in it well. 
Despite his uncooperative motives, in doing that Bob necessarily spends some of his own attention and contributes to the knowledge commons. He deserves to be rewarded, not so much for the production of $k'$, but for its implantation in $A'$. 
Because of the way authorship is defined in the MMM format, and because of the possibility of activity-based rewarding (cf \S\ref{rewarding}), Bob might generally expect better rewards for his publication of glue and his implanting efforts than for his publication of $k'$ \textit{per se}. Having documented glue between $k'$ and $A'$, it might  be in his best interests \textit{not} to refrain from also documenting glue  between $k$ and $k'$, and  between $A$ and $A'$. 

\subsection{Good Farmers and Good Farming}
There is a common situation that  lowers a \kpr{'s} need to consume collective attention, namely, when the \kpr{} cares about what  she produces and the knowledge she produces benefits  from the existence of prior knowledge that has already received attention. She may then capitalise on the  understanding that the community of  knowledge consumers already has of prior relevant knowledge in order to get the added value of her own work more efficiently noticed. Supporting \kpr{s} in this situation means facilitating the reuse of attention spent in the past in understanding prior knowledge. 
Good epistemic farming recognises that \KP{} works best through the exploitation of epistemic synergies\footnote{This can include contradiction. A new \pok{} $k'$ can benefit from the prior documentation of a contradicting \pok{} $k$. The two \poks{} $k'$ and $k$ can be connected in the MMM using a \mmmtype{differsFrom} edge labelled and/or annotated in such a way as to specify the contradiction. Through the \mmmtype{differsFrom} edge, the new \pok{} $k'$ may then benefit from some \kcr{s} having already paid attention to prior \pok{} $k$.} --
by relating new contributions  to pre-existing contributions (using meaningful links not prone to getting red-flagged and ignored), by demonstrating what the new contributions add to the record, how they build on or address prior knowledge.
Publishing knowledge no longer is enough like it is in academia. A  contribution must be findable and in the MMM system, the implantation task of ensuring that  it is, is incumbent upon the \kpr{}. And it isn't an administrative information retrieval task which merely demands of an  author  to tag 
their contribution $k$ with keywords  that non-specialists can manipulate. The task requires the author's
 attention and expertise in the field of $k$. This emphasis on field expertise negates the idea of a central communist system, and addresses 
Netting's remark: "\textit{Where tenure is poorly adapted to optimum land use \textbf{as seen by the cultivator}, productivity may be seriously curtailed.}" \cite{Netting}.

\subsection{House-Keeping}

Even if  the number of free-riding \kpr{s} were very small, the \CAR{} might still be overused. The rare free-riders could be  source of huge amounts of redundant free-riding contributions. And even without any free-riders at all,  responsible \kpr{s} still add \poks{} to the knowledge record. The record could still  grow at the risk of  shrinking collective attention. Maintenance of the record is needed: its size needs to be managed without depleting citizens' capacity to be well-informed. 
One way to do that 
is  to limit the addition of new \poks{}  by \kpr{s} {(cf \S\ref{seasonality})}.
Another way is house-cleaning. 
The MMM proposal offers solutions to deal with redundancy (cf \S\ref{redund}). It also offers obsoleting mechanisms that are not detailed in this paper \cite{article}. 
A \pok{}  that no \kcr{} wants to keep on their own land  spontaneously disappears from the MMM. 
A  \pok{} that is shared by some \kpr{s} and rejected by all \kcr{s} ends up occupying little space 
on the MMM, and  having little charge on the \CAR{}. The local house-keeping of \leties{}
by landowners  
constitutes a global house-keeping of the MMM commons. Every \kcr{'s} endeavour to control the use of their own attention (cf \S\ref{gatekeeping})  
\textit{directly} contributes to managing the use of the \CAR{}\footnote{In the attention economy, "\textit{To consume information, we must also be investors of our own attention portfolios.}" \cite{davenport2001attention}. The MMM allows this investment by individuals to have systematic positive externalities on the \CAR{}.}. 
Because duplicate \poks{} aggregate, and because \poks{} are atomic (as opposed to traditional documents like articles containing multiple atomic \poks{}), house-cleaning of MMM land is 
deep and rigorous (near exhaustive and fine-grained)\footnote{Compare with house-cleaning of a collection of overlapping documents copied in multiple folders of a file hierarchy. }, even in the absence of global regulations applying universally to the MMM commons. 
\smallskip

Deliberate centralised action is needed to ensure the persistence of old \poks{} that can make new \poks{} easier (less attention consuming) to understand -- after the old ones have fallen into general disinterest (they have disappeared from all \leties{})\footnotemark.  The identification of these \poks{} can rely on formal MMM based epistemic measures (cf \S\ref{topo}). 

\footnotetext{Note the asymmetry. Archiving knowledge can't rely on an undedicated crowd to select \poks{} to save despite the crowd's disinterest in them. On the contrary, mechanisms to filter out content undeserving of archival (and future attention), can leverage the crowd's disinterest. For instance 
\poks{} that have generated little to no discussion and disappeared very soon after their publication on the MMM might be excluded from systematic archival. }

\subsection{Enclosure}\label{enclosure}

We have discussed a desirable form of exclusion that protects a commons from overuse and from Hardin's tragedy. Enclosure denotes another  form of exclusion which deprives commoners of their right to enjoy a common resource. The resource is divided among private entities and the common disappears.  
Enclosure of physical land can result in commoners being locked out of their right to enjoy the land's resources  (e.g. wild cattle).  In the knowledge commons seen with the \KCV{}, the resources that individuals risk being locked out of are \poks{}. Enclosure means one or several central entities draw boundaries to define epistemic territories (cf echo chambers defined by  filter bubbles). The risk is mitigated on the MMM. In agreement with E. Ostrom's first principle for governing a commons \cite{OstromGov}, boundaries of MMM territories are clearly defined from the start.
The responsibility of what happens within those boundaries is already assigned to epistemic landowners (cf \S\ref{LETs}). And the MMM supports systematic epistemic bridging of territories (cf \S\ref{glue}). 
With the \KPV{}, the resource to consider is collective attention. The risk of being locked out of it entails  \textit{epistemic injustice} whereby some \kpr{s} are systematically ignored; see \cite{injustice}. 
The MMM solution provides gatekeeping. Gatekeeping applies to \poks{}, not to \kpr{s}. Also, it 
is implemented systematically on  \textit{local} epistemic territories, not on the global knowledge commons. And being epistemically driven rather than semantically driven, it
doesn't  apply to entire worldviews (cf \S\ref{echochambers}).
Arguably, the first of all knowledge problems to address, even before epistemic injustice, remains  to ensure that all individuals  have enough attention to produce and to consume \poks{} -- i.e., that they  not be excluded from enjoying their \textit{own} portion of the \CAR{}.

\section*{Conclusion} 

This article's \kpr{} view on the knowledge commons, emphasises epistemic attention. 
Why is epistemic attention important? Arguably because without it, one's reasoning is too shallow and unreliably well-informed to form an empowering understanding of the world \cite{injustice}. Davenport and Beck write:
"\textit{At one point, software magnates had the ambition to put "information at your fingertips." Now we've got it, and in vast quantities. But no-one will be informed by it, learn from it, act on it unless they've got some free attention to devote to the information.}" \cite{davenport2001attention}
Attention is the "\textit{limiting factor}" of the information economy \cite{davenport2001attention}, the missing link between the "\textit{bloomin' buzzin' confusion of the world around us [\ldots] and the decisions and actions necessary to make the world better}". 
\smallskip

With the technical MMM solution, an individual can exclude all (new) \poks{} from her \lety{}. 
This makes for a particularly free-minded but also empty-minded individual. Knowledge is an intrinsically  communal activity that brings people together. The knowledge commons has an advantage that other commons don't necessarily have: spontaneous \textit{disclosure}. Even the most stubborn minds participate in the knowledge economy. Everyone wants to consume some new information once in a while. And everyone has a thought to share once in a while that they want some other people to pay attention to. As long as (i) \kcr{s} and \kpr{s} have a say in how they spend their own attention, and (ii) there are \kcr{s} who want to consume new \poks{} and \kpr{s} who want to share new \poks{},
 there is a natural foundation for communal tenure. 
\smallskip

The  MMM solution focusses on supporting attention \textit{reuse}. 
It addresses both challenging sides of Davenport and Beck's attention equation \cite{davenport2001attention}: (1) how to get a hold of the attention of others, and (2) how to manage / allocate our own attention.
The MMM system is a technical solution to remedy or mitigate 
 typical knowledge commons problems such as: information overload (cf \hyperref[agg]{aggregation} and \hyperref[redund]{redundancy management}),  free-riders thoughtlessly relaying low quality content
(cf \hyperref[implant]{implantation} and \hyperref[rewarding]{possible reward based incentives}),
ignorance (cf \hyperref[topo]{formal epistemic measures}), and  low quality information itself (cf \hyperref[contimp]{continual annotation}).
The technical MMM solution pushes back the need for an evolving morality, narrowing it down to the following (hard) open questions:
How to define and gatekeep  "\textit{neutral}" public epistemic grounds (cf \S\ref{gatekeeping}, \S\ref{grounds})? How should such grounds be governed: communally and/or by a central public entity?   
What  kinds  of \poks{} should be systematically saved from the public's disinterest and archived for future generations?\footnote{As detailed in \cite{article}, the MMM proposal  offers the possibility of an  upgrade on the Internet Archive's precious \href{https://archive.org/web/}{Wayback Machine} through its support of a "structured epistemic time travel feature". What \poks{} are reachable by epistemic time travel depends on archiving choices.   }

\bibliography{bib.bib} 

\label{FSLastPage}

\end{document}